\newcommand\lsim{\mathrel{\rlap{\lower4pt\hbox{\hskip1pt$\sim$}}
    \raise1pt\hbox{$<$}}}
\newcommand\gsim{\mathrel{\rlap{\lower4pt\hbox{\hskip1pt$\sim$}}
    \raise1pt\hbox{$>$}}}

\documentstyle[prl,aps,epsf,floats,twocolumn]{revtex}

\begin{document}
\twocolumn[\hsize\textwidth\columnwidth\hsize\csname @twocolumnfalse\endcsname
\title{Cosmic string-seeded structure formation}

\author{P.P.~Avelino$^{1}$,
E.P.S.~Shellard$^{2}$, J.H.P.~Wu$^{2}$ and B.~Allen$^3$}
\address{$^{1}$Centro de Astrofisica, Universidade do Porto, Rua do Campo
Alegre 823,
4150 Porto, Portugal}
\address{$^{2}$Department of Applied Mathematics and Theoretical Physics,
University of Cambridge,\\
Silver Street, Cambridge CB3 9EW, U.K.\hspace*{2mm}
$^{3}$University of Wisconsin---Milwaukee, U.S.A.}

\maketitle

\begin{abstract}
We describe the results of high-resolution numerical simulations of
string-induced structure formation in open universes
and those with a non-zero cosmological constant.
For models with $\Gamma\,=\,\Omega h\,=\,0.1$--$0.2$ and a cold dark 
matter background, we show that the linear density fluctuation power
spectrum has both an amplitude at $8 h^{-1}$Mpc, $\sigma_8$, and an 
overall shape which are consistent within uncertainties with 
those currently inferred from galaxy surveys.
The cosmic string scenario with hot dark matter 
requires a strongly scale-dependent bias in order
to agree with observations.

\end{abstract}
\vskip .2in
]

\newpage
\noindent
{\bf A.\ Introduction}---In
this Letter we describe new results from an investigation of
cosmic string-seeded structure formation in hot and cold dark matter models.
The cosmic string scenario\cite{VilShe} predated inflation as
a realistic structure formation model, but it has proved computationally
much more challenging to make robust predictions with which
to confront observation.
The present paper relies on high resolution numerical simulations of a
cosmic string
network \cite{AS1} with a dynamic range extending from before the
matter-radiation transition through to deep in the matter era
(developing on previous work \cite{AveShe}).
We calculate
the linear power spectrum of density perturbations ${\cal P}(k)$ induced by
the strings in flat models with and without a cosmological constant,
and we then extrapolate to open cosmologies.
This work represents
a considerable quantitative advance by
incorporating
important aspects of the relevant physics not included in previous
treatments. 

In the first instance,
we consider density perturbations about a flat
Friedmann-Robertson-Walker (FRW) model
with a cosmological constant $\Lambda$
and which are causally
sourced by an evolving string network with energy-momentum tensor
$\Theta_{\alpha\beta}({\bf x}, \eta)$.  In the synchronous gauge, the
linear evolution equations of the radiation and cold dark matter (CDM)
perturbations,
$\delta_{\rm r}$ and $\delta_{\rm c}$ respectively, are given
by (modified from \cite{VeeSte})
\begin{eqnarray}
\ddot \delta_{\rm c} + {\dot a \over a} \dot \delta_{\rm c} -
{3 \over 2}\Big({\dot a \over a}\Big)^2 \, \left({a
\delta_{\rm c} + 2 a_{\rm eq} \delta_{\rm r}
\over a + a_{\rm eq}
+ \frac{\Omega_{\Lambda}}{\Omega_{\rm c}}\frac{a^4}{a_0^3}}
\right)
= 4 \pi G \Theta_+,
\label{two} \\
\ddot \delta_{\rm r} - {1 \over 3} \nabla^2 \delta_{\rm r}
- {4 \over 3}
\ddot \delta_{\rm c} = 0,~~~~~~~~~~
\label{three}
\end{eqnarray}
where $\Theta_+ =  \Theta_{00} +\Theta_{ii}$, $a$ is the scale factor,
the subscript ``eq'' denotes the epoch of radiation-matter density equality,
``0'' denotes the epoch today,
a dot represents a derivative with
respect to conformal time $\eta$,
$\Omega_{\rm c}=
  8\pi G \rho_{\rm{c0}}/3H_0^2$ and
$\Omega_\Lambda=\Lambda/3H_0^2$.
It proves useful to split
these linear perturbations into
initial (I) and subsequent (S)
parts\cite{VeeSte},
$
\delta_N ({\bf x}, \eta) = \delta_N^{\rm I}({\bf x}, \eta)
+ \delta_N^{\rm S}({\bf x}, \eta)\,,$ where $N={\rm c, r}$.
The initial perturbations $\delta^{\rm I}({\bf x}, \eta)$ depend on the string
configuration at some early time $\eta_{\rm i}$, because the
formation of strings creates underdensities in the initially homogeneous
background out of which they are carved.
The subsequent perturbations $\delta^{\rm S}({\bf x}, \eta)$ are those
which are generated actively by the strings themselves
for $\eta>\eta_{\rm i}$.
Because strings induce isocurvature perturbations,
$\delta^{\rm I}({\bf x}, \eta)$
must compensate  $\delta^{\rm S}({\bf x}, \eta)$ on comoving scales
$|{\bf x}-{\bf x}'|>\eta$ to prevent acausal fluctuation growth on
superhorizon scales.

The system of equations (\ref{two},\ref{three}) can be solved for the
subsequent perturbations  $\delta^{\rm S}({\bf x},\eta)$,
with initial conditions
$\delta_{\rm c}^{\rm S}\!=\!\delta_{\rm r}^{\rm S}\!=\!0$ and
${\dot \delta}_{\rm c}^{\rm S}\!=\!{\dot \delta}_{\rm r}^{\rm S}\!=\!0$
at $\eta\!=\!\eta_{\rm i}$, by using a
discretized version of the integral equation with Green's functions:
\begin{equation}
  \delta^{\rm S}_N({\bf x},\eta) = 4 \pi G \!\!\int_{\eta_{\rm i}}^{\eta} \!\!\! d\eta'
  \int \!\!d^3x'\: {\cal G}_N (X;\eta,\eta') \Theta_{+}({\bf x'},\eta'),
\label{delta_S_N}
\end{equation}
where $X\!=\!|{\bf x}-{\bf x'}|$. The Green's functions in Fourier space
can be calculated numerically by solving the homogeneous version of 
(\ref{two},\ref{three}) with initial conditions at $\eta\!=\!\eta'$:
$\dot{\widetilde{\cal G}}_{\rm c}\!=\!3\dot{\widetilde{\cal G}}_{\rm r}/4\!=\!1$
and
$\widetilde{\cal G}_{\rm c}\!=\!\widetilde{\cal G}_{\rm r}\!=\!0$
($\widetilde{\cal G}_N\!=\!0$ for $\eta\!<\!\eta'$).

The
subsequent perturbations $\delta^{\rm S}({\bf x},\eta)$ 
are dynamically sourced by moving local strings with
spacetime trajectories we can represent as $x^\mu_{\rm s} = (\eta, {\bf
x}_{\rm s}(\sigma,\eta))$, where $\sigma$ is a spacelike parameter labelling
points along the string (a prime represents a derivative
with respect to $\sigma$).  The stress energy tensor of the string source
is then given by\cite{VilShe}
\begin{equation}
\Theta_{\mu \nu}({\bf x}, \eta) = \mu
\int d \sigma
(\epsilon {\dot x_{\rm s}}^{\mu} {\dot x}_{\rm s}^{\nu} -
 \epsilon^{-1}{x_{\rm s}'}^{\mu} {x_{\rm s}'}^{\nu})
\delta^3 ({\bf x} - {\bf x}_{\rm s}),
\end{equation}
where $\mu$ is the string linear energy density,
$\epsilon = [{\bf x_{\rm s}'}^2 / (1 - {\bf {\dot x_{\rm s}}}^2)]^{1/2}$,
and we have also assumed that ${\dot {\bf x}_{\rm s}} \cdot  {{\bf x_{\rm
s}}'}=0$.
In this case, it is
straightforward to compute $\Theta_+$ in (\ref{two}) as
\begin{equation}
\Theta_+({\bf x},\eta)=
\Theta_{00} +\Theta_{ii}=
 2 \mu \int d \sigma \epsilon
{\bf {\dot x_{\rm s}}}^2 \delta^3 ({\bf x} - {\bf x_{\rm s}}).
\label{theta_plus}
\end{equation}
The stress energy $\Theta_{\mu\nu}$ was calculated directly from high resolution
string network simulations\cite{AS1}.  Dynamical ranges exceeding 100 in 
conformal time (redshifts up to 1000) were achievable because of a 
`point-joining' algorithm maintaining fixed comoving resolution\cite{Rob}
and parallelization.

\noindent
{\bf B.\ Approximation schemes}---It
is a very substantial numerical challenge to evolve the
initial and subsequent perturbations induced by cosmic strings
such that they accurately cancel on superhorizon scales by the present
day $\eta_0$.
For the
large dynamic range required for the present study, we have
by necessity adopted the `compensation factor approximation' suggested in
a semi-analytic context in ref.~\cite{AScdm}.  To implement this,
we accurately evolved the long string network numerically---the dominant
active source term---and then multiplied the Fourier transform
of the resulting stress energy $\widetilde\Theta_{+}(k,\eta)$ by
a cut-off function
${\widetilde F}(k,\eta) = [1+(k_{\rm c}/k)^2]^{-1}$.
This results in the correct $k^4$ fall-off in the power spectrum at large
wavelengths above the compensation scale $k_{\rm c}^{-1}\sim \eta$.
The efficacy
of this approximation has been demonstrated
by studying multifluid compensation backreaction effects in
ref.~\cite{painless}.  
For the present study we have adopted the
analytic fit for $k_{\rm c}(\eta)$ presented in ref.~\cite{painless},
which smoothly interpolates from $k_{\rm c}
=\sqrt{6}\eta^{-1}$ in the radiation era to $k_{\rm c}=\sqrt{18}\eta^{-1}$
in the matter era.
The quantitative implementation of compensation is a subtle issue
and a key uncertainty in all work to date on gauged
cosmic strings.  We note, however, that our results are relatively insensitive
to the choice of $k_{\rm c}$, especially in open and $\Lambda$-models (e.g.
a large factor of 2 increase in $k_{\rm c}$ causes only about a 20\% decrease 
in the power spectrum at $k \approx 0.15 h^{-1}$Mpc for a 
flat $\Omega_\Lambda=0.8$ model).

In order to study the formation of structures with cosmic strings
in hot dark matter (HDM) models, we use
a reasonably accurate alternative to much
more elaborate calculations using the
collisionless Boltzmann equation.
We simply multiply the Fourier transform of the string source term
$\widetilde\Theta_+(k, \eta)$
by a damping factor
${\widetilde G}(k,\eta)={[{{1+(0.435 k
D(\eta))}^{2.03}}]}^{-4.43}$
\cite{AShdm}.
Here, $D(\eta)$ is the comoving damping length that a
neutrino with velocity $T_\nu/m_\nu$ can travel from the time $\eta$ onwards.
The factor ${\widetilde G}(k,\eta)$ is a fit to numerical calculations of
the transfer function
of a Fermi-Dirac distribution of non-relativistic neutrinos and accounts for
the damping of small-scale perturbations due to neutrino
free-streaming\cite{AShdm}.
We calculated $D(\eta)$ numerically and found
an excellent fit to our results
for $T_{\nu 0}=1.6914 \times 10^{-13} \, {\rm GeV}$ and $m_\nu=91.5 \,
\Omega h^2 \, {\rm eV}$ with $
  D(\eta)=\frac{1}{20}\log{\left[{(5\eta_{\rm eq}+\eta})/\eta\right]}
$.

The other key difficulty confronting defect simulations is their
limited dynamic range.  At any one time, an evolving string network sources
significant power over a
lengthscale range which exceeds an order of magnitude.
However, we can employ a semi-analytic model to compensate
for this missing power\cite{AScdm}, which proves to be fairly
accurate in the scaling regimes away from the matter-radiation transition.
The procedure is essentially to square the expression (\ref{delta_S_N}) in
Fourier space
to obtain the power spectrum ${\cal P}(k)$.  This becomes a Green's function
integral over the unequal time correlators $\langle \Theta_+({\bf k},\eta)
\Theta_+({\bf -k},\eta')\rangle$, which are subsumed in a scale-invariant
`string structure function' ${\cal F}({\bf k},\eta)$, that is, the 
power spectrum is given by\cite{AScdm}:
\begin{equation}
  \label{semi-ana}
  {\cal P}(k)=16\pi^2 G^2 \mu^2 \int_{\eta_{\rm i}}^{\eta_0}
  |{\cal G}_{ N}(k;\eta_0,\eta)|^2 {\cal F}(k, \eta) d\eta\,.
\end{equation}
In practice, we obtained the structure function ${\cal F}({\bf k},\eta)$
phenomenologically by fitting its shape and amplitude to simulations of
limited dynamic range deep in the matter and radiation eras.  We were then
able to use an interpolation based on the actual string density during
the transition era to provide a good fit to the simulation power
spectrum for any given dynamic range $\eta_{\rm i} \rightarrow \eta$.

\noindent
{\bf C.\ Open and $\Lambda$-cosmologies}---The possibility that the universe
is open or has a cosmological constant is now favoured by a number of 
oberservations, so
it is natural to explore the string-induced spectrum in these two regimes.
For the $\Lambda$-models $\Omega_{\rm c} + \Omega_\Lambda = 1$,
we have taken the weak dependence for the
COBE normalisation of
$G\mu\propto \Omega_{\rm c}^{-0.05}$ suggested in ref.~\cite{9708057}
for (\ref{delta_S_N},\ref{semi-ana}).
For the open models $\Omega_{\rm c}<1$ and $\Lambda=0$,
we simply rescale the simulated
spectrum from a flat universe with $\Lambda=0$ in the
following way (adapted from \cite{9708057}):
\begin{equation}
  \label{Sopen}
  S(k,h,\Omega_{\rm c}) =
  S(k,1,1) \cdot
  \Omega_{\rm c}^2 {h}^4 \cdot f^2(\Omega_{\rm c}) \cdot
  g^2(\Omega_{\rm c}),
\end{equation}
where $k$ is in units of $\Omega_{\rm c}h^2 \, {\rm Mpc}^{-1}$ and
$f(\Omega_{\rm c})=\Omega_{\rm c}^{-0.3}$
reflects the COBE normalisation of $G\mu$ \cite{9708057}. The last factor,
$g(\Omega_{\rm c})=
  2.5\Omega_{\rm c}
  /({1 + \Omega_{\rm c}/{2} + \Omega_{\rm c}^{4/7}})$,
gives the total suppression of linear growth for density perturbations
in an open universe
relative to an $\Omega_{\rm c}\!=\!1$ and $ \Omega_\Lambda\!=\!0$
universe\cite{Carroll_Einstein}.
Finally, the second factor $\Omega_{\rm c}^2h^4$ in (\ref{Sopen}) 
arises naturally from the
normalization of the Green's functions in (\ref{two},\ref{three}).
We have verified that a similar analytic rescaling from an
$\Omega_{\rm c}=1, \Omega_\Lambda=0$ model to 
$\Omega_{\rm c} + \Omega_\Lambda=1$ models
agrees very accurately with the solutions of (\ref{two},\ref{three})
obtained from simulations.
This also helps to justify the extrapolation still required for open models,
although some uncertainty remains concerning the COBE normalization.

\begin{figure}
\centering 
\leavevmode\epsfxsize=8cm \epsfbox{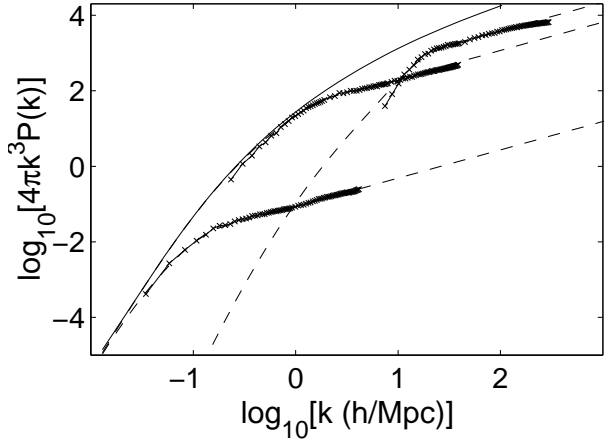}\\
\caption[]
{Comparison of the string-induced simulation power
  spectrum and the semi-analytic fit (\ref{semi-ana}).
  The top-right, central, and bottom-left solid lines with crosses are the
  simulation results in the deep radiation, transition,
  and deep matter eras respectively.
  The dashed lines are the semi-analytic fits corresponding to the same
  dynamical ranges of the simulations.
  The solid line is the semi-analytic model over the full dynamic range
  from $\eta_{\rm i}=0$ to today.
  }
\label{open_fig1}
\end{figure}

\noindent
{\bf D.\ Results and discussion}---In
figure \ref{open_fig1} we plot the Cosmic Microwave Background (CMB) normalized
(i.e. $G\mu_6 = G\mu \times 10^6 = 1.7$\cite{ACDKSS})
linear power spectrum
induced by cosmic strings in an $\Omega_{\rm c}=1$ CDM cosmology
with $h=0.7$.
The central set of numerical points
was sourced by string network simulations beginning at $\eta=0.4\eta_{\rm
eq}$ which were continued for 1318 expansion times
(from redshift $z_{\rm i}\approx 31700$ to $z_{\rm f}\approx 23$).
String simulations were always ended before the horizon grew to 
half the simulation box-size. These had a string sampling resolution
at least four times higher than the structure formation grids, which had
up to $256^3$ points with overall physical scales ranging 
from 4--100$h^{-1}$Mpc.
Given the dynamic range limitations, we
have also plotted the semi-analytic fit (\ref{semi-ana}) over the full
range of wavenumbers, illustrating the good agreement with our numerical
results.  This was also apparent in
HDM simulations, so we have considerable confidence
that the semi-analytic model provides a good approximation to the shape and amplitude of
the string simulation power spectrum.  These results are also qualitatively 
consistent with the semi-analytic results of ref.~\cite{AScdm} 
and also with unpublished matter era simulations \cite{Albert}.

\begin{figure}[t]
\centering 
\leavevmode\epsfxsize=8cm \epsfbox{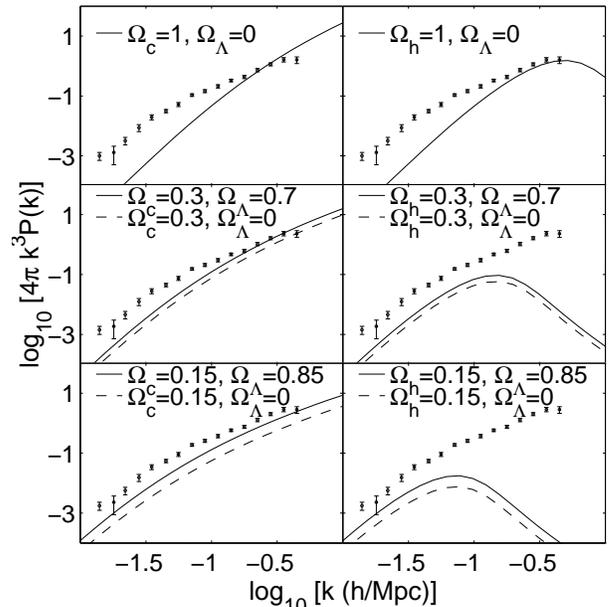}\\ 
\caption[]
{The CMB normalized linear spectra induced by 
  cosmic strings in CDM (left) and HDM (right)
  models for different background cosmologies.        
  The solid lines are the flat $\Lambda$-models;
  the dashed lines are the open models.
  Here, we use $h=0.7$ and
  the data points with error bars are the linear spectrum
  reconstructed from observations \cite{PD}.
  }
\label{open_fig2}
\end{figure}

In figure \ref{open_fig2}, we make a comparison between our
CDM and HDM string power spectra and the
observational results inferred from galaxy surveys \cite{PD}, in each of five
different background cosmologies:
(I) $\Omega_{\rm c,h} =1$, $\Omega_{\rm \Lambda} =0 $,
(II) $\Omega_{\rm c,h} =0.3$, $\Omega_{\rm \Lambda} =0.7 $,
(III) $\Omega_{\rm c,h} =0.3$, $\Omega_{\rm \Lambda} =0 $,
(IV) $\Omega_{\rm c,h} =0.15$, $\Omega_{\rm \Lambda} =0.85 $,
and (V) $\Omega_{\rm c,h} =0.15$, $\Omega_{\rm \Lambda} =0 $.

Consider first the $\Omega_{\rm c}=1$ CDM
model.  We calculated the standard deviation of density perturbations
$\sigma_8$ by convolving with a spherical window of radius $8 \, h^{-1} \,$Mpc
to find
$\sigma_{\rm 8(sim)}(h=0.5)=0.32 G\mu_6$,
$\sigma_{\rm 8(sim)}(h=0.7)=0.39 G\mu_6$ and
$\sigma_{\rm 8(sim)}(h=1.0)=0.47 G\mu_6$.
A comparison with the observational data
points shows that strings appear to induce an excess of
 small-scale power and
a shortage of large-scale power, that is, the $\Omega_{\rm c}=1$ string
model requires a significant scale-dependent bias.
This is not necessarily a fatal flaw on small scales
because, as the corresponding HDM spectrum indicates, such excess power
can be readily eliminated in a mixed dark matter model.  However,
the problem is less tractable on large scales where
biases up to $\sigma_{\rm 100(obs)}/\sigma_{\rm 100(sim)}\approx 3.9$ 
around 100$h^{-1}$Mpc
might be inferred from the data points (using
$G\mu_6= 1.7$ and $h=0.7$).
Should we, therefore, rule out string models on this basis\cite{against}?
Although the 
$\Omega_{\rm c}=1$ spectrum looks unattractive, there are three
important mitigating factors.  First, the present observational
determination of the
power spectrum around 100$h^{-1}$Mpc is very uncertain.   Secondly, the
immediate nonlinearity of
string wakes means that strong biasing mechanisms might operate
on large scales.
Finally, unlike inflation,
defect models have never been wedded to an $\Omega=1$ cosmology.

[b]
\begin{figure}
\centering 
\leavevmode\epsfxsize=8.5cm \epsfbox{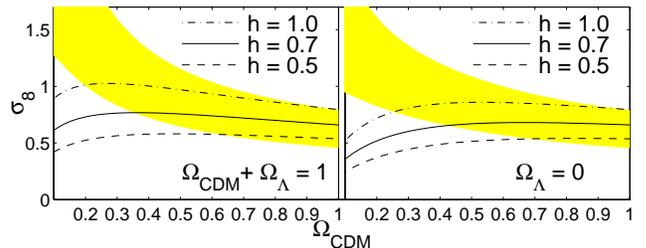}\\ 
  \caption[]
  {A comparison of the observationally inferred mass fluctuation 
    $\sigma_{\rm 8(obs)}$ with that 
    induced by cosmic strings in our simulations, $\sigma_{\rm 8(sim)}$.
    $\sigma_{\rm 8(obs)}$ is shown as the shaded area \cite{VL}, while
    $\sigma_{\rm 8(sim)}$ is plotted as
    dot-dashed ($h=1.0$),
    solid ($h=0.7$) and
    dashed($h=0.5$) lines. 
    }
  \label{open_fig3}
\end{figure}
We can observe from figure \ref{open_fig2}, that for open or
$\Lambda$-cosmologies
with $\Omega_{\rm c}\approx 0.1$--$0.3$, the string $+$ CDM power
spectrum is much more encouraging. We find that the bias on large scales
is always reasonably close to unity and, overall, it is much less 
scale-dependent.
For example, over the full range of lengthscales in model IV  
($\Omega_{\rm c} =0.15$,
$\Omega_{\rm \Lambda} =0.85$),
the relative bias remains
$\sigma_{\rm 100(obs)}/\sigma_{\rm 100(sim)}\approx 1.4 \pm 0.2$
at 100$h^{-1}$Mpc.
In figure \ref{open_fig3},
the value of $\sigma_{\rm 8(sim)}$ induced in our simulations
with the CMB normalized $G\mu_6$ \cite{9708057,ACDKSS} is
compared with the observationally inferred 
$\sigma_{\rm 8(obs)}$ \cite{VL}
for the full gamut of open and $\Lambda$-models.
We can see from figure \ref{open_fig3}, that
$\sigma_{\rm 8(obs)}/\sigma_{\rm 8(sim)}(\Omega_{\rm c}$$=$$1)\approx
0.79 \pm 0.21,\; 0.95 \pm 0.25,\; 1.17 \pm 0.31$
for $h=1.0,\, 0.7,\, 0.5$ respectively. 
When $h=0.7$, $\sigma_{\rm 8(sim)}$ matches $\sigma_{\rm 8(obs)}$ within
the uncertainties
for flat $\Lambda$-models when $\Omega_{\rm c}\gsim 0.35$
and for open models when $\Omega_{\rm c}\gsim 0.4$, while for both cases
the ratio $\sigma_{\rm 8(obs)}/\sigma_{\rm 8(sim)}\lsim 2$ for all 
$\Omega_{\rm c}\gsim 0.1$.
Combining these results with an analysis similar to 
figure \ref{open_fig2}, we found that the best string models lie in the 
range $\Gamma=\Omega h=0.1$--$0.2$, producing both an acceptable 
$\sigma_{\rm 8(sim)}$ and power spectrum shape. 
Hence, an open or $\Lambda$-cosmology in the context of string $+$ CDM model
certainly merits a more detailed nonlinear study.
These conclusions are in qualitative agreement with semi-analytic results
\cite{9708057,Ferreira} 
and those based on a phenomenological string model \cite{ABR2}.

As for the HDM results, the comparison with observation
seems to require a strongly scale-dependent bias for any choice of the
cosmological parameters (models I--V). However,
the lack of small scale power may be partially overcome
if baryons are properly included in the analysis. Further investigation
using a hydrodynamical code will be required to determine
whether galaxies can form early enough.

A key feature of all these string-induced power spectra is the influence
of the slow relaxation to the matter era string density from the much
higher radiation string density, which has an effective
structure function ${\cal F}(k,\eta)$ in (\ref{semi-ana})
with approximately 2.5 times more power than the matter era version.
Even by recombination in an $\Omega_{\rm c}=1$ cosmology, the string
density is more than twice its asymptotic matter era value to which we
normalize on COBE scales.  This implies that the string model provides
higher than expected large-scale power around 100$h^{-1}$Mpc and below.
Interestingly, this can also be expected to produce a significant
Doppler-like peak on small angle CMB scales, an effect noted in
ref.~\cite{ACDKSS} but not observed because only matter era 
strings were employed.
Recent work in ref.~\cite{ABR2} confirms that such Doppler-like features can
result from significant non-scaling effects during the transition
era.

Finally, we comment on the fact that
the key uncertainties affecting these calculations primarily influence
the amplitude of the string power spectrum, rather
than its overall shape which appears to be a more robust feature.
These uncertainties mainly result from the compensation approximation
(mentioned previously),
the COBE normalization of the string energy density \cite{ACDKSS,ACSSV},
the analytic approximation to the Green's functions,
and systematic errors \cite{AWSA}.  Combining our best estimates of these
uncertainties gives an approximate factor of 2 uncertainty in the power 
spectrum amplitude for 
$\Omega_{\rm c}=1$ and $\Lambda$-models. The extrapolations required 
for open models with $\Gamma \approx 0.15$ increase this uncertainty to at 
least a factor of 3 overall, but we will discuss this at length 
elsewhere~\cite{AWSA}.

\noindent
{\bf E.\ Conclusion}---We have described the results of high-resolution
numerical simulations of structure formation
seeded by a cosmic string network with a large dynamical range taking into
account, for the first time,  modifications 
due to the radiation-matter transition.
Our results show that
for $\Gamma\!=\!\Omega h\!=\!0.1$--$0.2$
both $\sigma_8$ and
the power spectrum shape of cosmic string-induced CDM fluctuations 
agree satisfactorily with observations.
In particular,
the generalization to open or $\Lambda$-models tends to remove the 
excess small-scale power found in cosmic string models 
with $\Omega_{\rm c}\!=\!1$ and $\Omega_\Lambda\!=\!0$, while also 
bolstering the large-scale power.
The HDM power spectrum requires a strongly scale-dependent bias
either on small or large scales,
but we note that a high baryon fraction may help to increase
small-scale power.
We conclude that
the picture which emerges for particular cosmic string models
seems encouraging and certainly deserves further study \cite{AWSA}.

We thank Carlos Martins, Robert Caldwell,
Richard Battye, and Pedro Viana for useful conversations.
PPA is funded by JNICT 
(PRAXIS XXI/BPD/9901/96).
JHPW is funded by CVCP 
(ORS/96009158) and by the Cambridge Overseas Trust (UK).
This work was performed on the COSMOS Origin2000 supercomputer
which is supported by Silicon Graphics, 
HEFCE and PPARC.



\end{document}